\documentclass[conference]{IEEEtran}
\IEEEoverridecommandlockouts
\usepackage{cite}
\usepackage{amsmath,amssymb,amsfonts}
\usepackage{algorithmic}
\usepackage{graphicx}
\usepackage{textcomp}
\usepackage{mathtools}
\usepackage{listings}
\usepackage[table,x11names]{xcolor}
\def\BibTeX{{\rm B\kern-.05em{\sc i\kern-.025em b}\kern-.08em
    T\kern-.1667em\lower.7ex\hbox{E}\kern-.125emX}}
\usepackage{soul}

\newcommand \row[1] {\texttt{row} ({#1}_{n-1}, \; #1_{n-2}\dots,\; #1_{0})}

\newcommand \rrow[1] {\texttt{row} ({#1}, \; #1\dots,\; #1)}
\newcommand \shrow[1] {\texttt{row} ({#1}_{n-1}, \dots,\; #1_{0})}
\usepackage{caption}
\usepackage{subcaption}
\usepackage{array}
\usepackage{arydshln}
\usepackage{tikz}
\usetikzlibrary{shapes,arrows,quotes}
\usepackage{pgfplots}
\pgfplotsset{compat=1.18, width=7.5cm}
\definecolor{scgreen}{HTML}{7fc97f}
\definecolor{scorange}{HTML}{fdc086}
\definecolor{scpurple}{HTML}{beaed4}
\definecolor{scred}{HTML}{e41a1c}
\definecolor{scdgreen}{HTML}{4daf4a}
\definecolor{scyellow}{HTML}{ffff99}

\begin{document}

\title{Multiplier Optimization via E-Graph Rewriting
}


\author{
\IEEEauthorblockN{Andy Wanna$^1$, Samuel Coward$^{1,2}$, Theo Drane$^2$, George A.~Constantinides$^1$ and Milo\v{s} D. Ercegovac$^3$}
\IEEEauthorblockA{$^1$ Imperial College London, $^2$ Intel Corporation, $^3$ University of California, Los Angeles\\
Email: \{andy.wanna20, s.coward21, g.constantinides\}@imperial.ac.uk, theo.drane@intel.com, milos@cs.ucla.edu}
}

\maketitle

\begin{abstract}
Multiplier circuits account for significant resource usage in datapath-dominated circuit designs, and RTL designers continue to build bespoke hand-crafted multiplication arrays for their particular application. The construction of an optimized multiplier presents trade-offs between pre-processing to generate a smaller array and array reduction. A data structure known as an e-graph has recently been applied to datapath optimization, where the e-graph's ability to efficiently explore trade-offs has been shown to be crucial. We propose an e-graph based rewriting framework to construct optimized multiplier circuits. Such a framework can express alternative multiplier representations and generate customized circuit designs. We demonstrate that the proposed tool, which we call OptiMult, can reduce the latency of a squarer by up to 46\% and reduce the latency of a standard multiplier by up to 9\% when compared against logic synthesis instantiated components.
\end{abstract}


\section{Introduction}
Multiplication circuits consist of three stages: array creation, array reduction and a final carry-propagate adder. We re-consider these boundaries, blurring
the line between array reduction and carry-propagation. The most common array creation approaches deploy AND arrays or Booth encoding at various radices~\cite{Ercegovac2004DigitalArithmetic, Koren2018ComputerAlgorithms}. Array reduction is typically achieved via compressor cells that sum columns of bits~\cite{Ercegovac2004DigitalArithmetic,Stenzel1977AScheme}, or reduce $n$ rows to $m$ rows, where $m<n$. Commonly used reduction schemes are the Wallace tree~\cite{Wallace1964AMultiplier} and Dadda tree~\cite{Dadda1965SomeMultipliers}. High-level and logic synthesis tools automatically replace $a*b$ in source code with optimized components, but rely on fixed architectures and logic synthesis gate-level optimization. Automated techniques for compressor tree synthesis have explored both heuristic based search methods and integer linear programming~\cite{Kumm2018AdvancedFPGAs,Parandeh-Afshar2008ImprovingProgramming}. 

We propose an automated rewriting framework helping RTL designers to explore bespoke multiplier implementations. The framework is based on the e(quivalence)-graph data structure that efficiently explores equivalent designs by clustering equivalent sub-expressions into e(quivalence)-classes. This clustering into e-classes captures the notion that there are many equivalent implementations of a given sub-graph in a dataflow graph. The e-graph is grown via constructive rewrite application, meaning that the left-hand side of the rewrite is retained rather than replaced. E-graphs have recently been successfully applied to datapath optimization~\cite{Coward2022AutomaticE-Graphs,Ustun2022IMpress:HLS,Coward2023AutomatingE-Graphs}. 

Typical e-graph optimization tools explore a single e-graph, however this is not scalable for many applications. Prior work has proposed sketch-guided equality saturation~\cite{Koehler2021Sketch-GuidedPrograms}, where intermediate sketches enable e-graph contraction and re-initialization. We modify such an approach by encoding intermediate e-graph optimization phases in a sequence of cost models. We have developed a multiplication circuit optimizer, OptiMult, based on the extensible e-graph library, \texttt{egg}~\cite{Willsey2021Egg:Saturation}, targeting minimal latency circuits. This paper contains the following novel contributions:
\begin{itemize}
    \item application of e-graphs to gate-level component design,
    \item a multi-level and iterative e-graph optimization method,
    \item a dual representation of AND arrays enabling the application of column-wise and row-wise optimizations,
    \item expression of multiplier optimizations as a set of local equivalence-preserving rewrites.
\end{itemize}

\section{Methodology}
The approach is separated into three key contributions. First we describe the representations that make multiplier design amenable to rewriting. Next we describe how multiplier optimization can be decomposed into a collection of local equivalence preserving rewrites. Lastly, we address scalability, taking inspiration from compilers~\cite{Cooper2011EngineeringEdition}, and separating optimization into a set of passes. 

\subsection{Multiplier Representation} \label{sec:representation}
\begin{figure*}
     \centering
     \begin{subfigure}[t]{0.62\textwidth}
         \centering
            \includegraphics[scale=0.35]{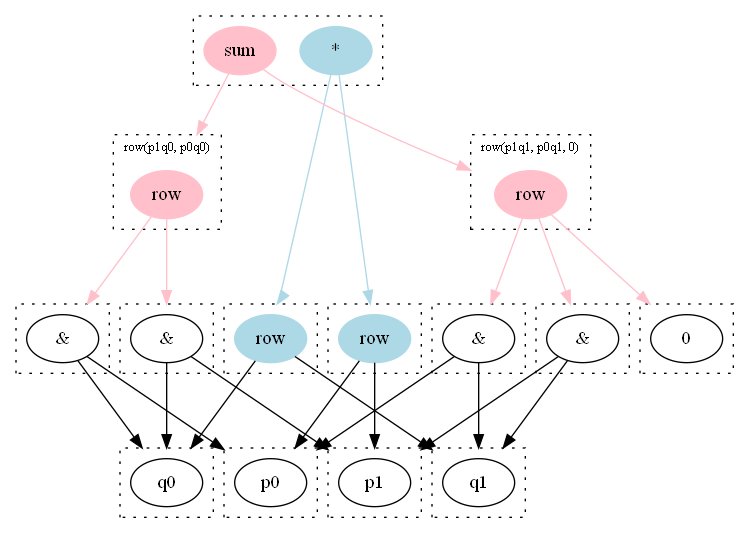}
         \caption{Applying the initial mapping from the multiplier representation (blue), the e-graph contains an equivalent sum of rows representation (pink).}
         \label{fig:stage_1}
     \end{subfigure}
     \hfill
     \begin{subfigure}[b]{0.33\textwidth}
         \centering
    \begin{tabular}{c|cccc}
         \rowcolor{pink!30}\cellcolor{white}&\texttt{row}&           & p1q0 & p0q0 \\
         \rowcolor{pink!50}\cellcolor{white}\cellcolor{white}\texttt{sum}&\texttt{row}& p1q1 & p0q1 & 0  \\
         \hline
    \end{tabular}
    \caption{Sum of rows representation (pink).}
    \label{tab:sum_of_rows}
    
    \newcolumntype{g}{>{\columncolor{orange!20}}c}
    \newcolumntype{h}{>{\columncolor{orange!30}}c}
    \newcolumntype{k}{>{\columncolor{orange!10}}c}
    \begin{tabular}{cc|k|g|h|}
    \\
    \\
    \\
        && \texttt{sum}&\texttt{sum}&\texttt{sum}\\
         &         & & p1q0 & p0q0 \\
        &\texttt{row} &p1q1 & p0q1 & \\
        \hline
    \end{tabular}
    \caption{Row of sums representation (orange).}
    \label{tab:row_of_sums}
     \end{subfigure}
     \hfill
     \begin{subfigure}[b]{0.65\textwidth}
         \centering
            \includegraphics[scale=0.35]{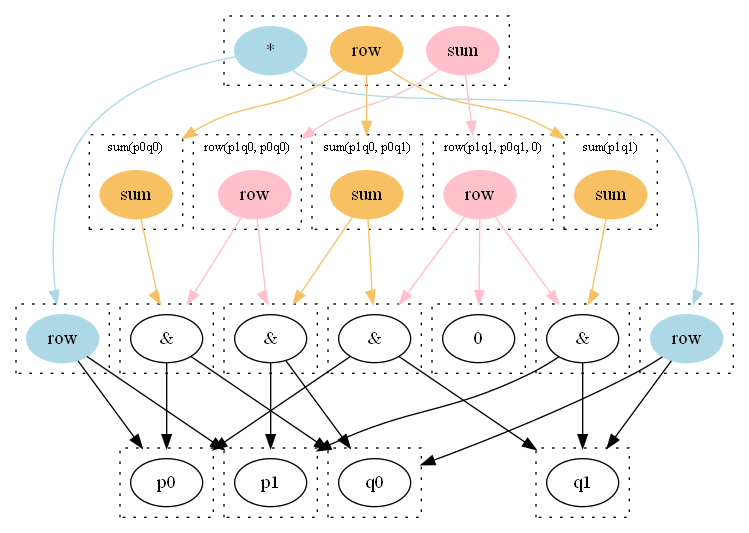}
         \caption{After applying the ``sum-of-rows'' rewrite, the e-graph contains an equivalent row of sums representation (orange).}
         \label{fig:stage_2}
     \end{subfigure}

        \caption{Stages of e-graph growth as rewrites are applied for a 2-bit multiplier circuit producing $\{p_1,p_0\}*\{q_1,q_0\}$. The dashed boxes denote e-class boundaries. The top e-class grows to contain three nodes, each defining an equivalent multiplier representation. Class annotations are added to describe the operation, where $ab$ is short for $a\,\&\,b$.}
        \label{fig:e-graph-stages}
\end{figure*}

In this work an input multiplier expressed as a simple $*$ operator, is initially mapped to an AND array. The AND array is represented as an e-graph, where each node is either an operator connected by edges to its children or a variable/constant. The internal representation supports the basic logical operators, addition of bits and several specialized operators. The \texttt{row} operator is equivalent to a concatenation of its children. The \texttt{sum} operator computes the addition of \texttt{row}s, where each \texttt{row} may represent a multi-bit number. Figure~\ref{fig:e-graph-stages} depicts these different representations in both an e-graph and as an array for a 2-bit multiplier, $\{p_1,p_0\} * \{q_1,q_0\}$, where $p_i$ and $q_i$ are single bits. The rewrites described in Section~\ref{sec:rewrites} encode the transitions between these two representations and highlight the benefit of maintaining multiple representations. 

In addition to the \texttt{row} and \texttt{sum} operators we also include operators for the commonly-used compressor cells. Specifically, we include a Full-Adder Carry ($FA_c$) and Sum ($FA_s$) each taking three operand bits, as well as a Half-Adder Carry ($HA_c$) and Sum ($HA_s$) which take two operand bits. These operators allow OptiMult to efficiently explore array reduction schemes at a higher abstraction level, rather than decomposing these operators into their gate-level implementations. The decreasing level of abstraction optimizations are discussed in Sections~\ref{sec:rewrites} and~\ref{sec:schedule}.

\subsection{Rewriting for Multiplier Optimization}\label{sec:rewrites}
\begin{table*}
    \centering
    \caption{Local equivalence preserving rewrites. Rewrites marked with $\dagger$ are dynamic rewrites, meaning these represent example instances as their right-hand side is constructed at runtime. All rewrites match against arbitrary row length $n$ or $m$. The ``divide-and-conquer'' rewrite is a pre-processing rewrite applied for larger multipliers. It is expressed here in terms of $*$ operators which can each be converted to an AND array as shown in Figure~\ref{fig:divide_and_conq}.}
    \label{tab:rewrites}
    \setlength\extrarowheight{2pt}
    \rowcolors{2}{white}{gray!25}
    \begin{tabular}{|c|l|c|c|}
        \hline
         Phase & Rewrite Name & Left & Right \\
         \hline
         1 & Place Half-Adder          & $a+b$                     & $\texttt{row} (HA_c(a,b), \; HA_s(a,b))$ \\
         1 & Place Full-Adder          & $a+b+c$                   & $\texttt{row} (FA_c(a,b,c), \; FA_s(a,b,c))$ \\
         1 & add-same                  & $a+a$                     & $\texttt{row} (a, \; 0)$ \\
         1 & row-add    $\dagger$    & $a + \row{b}$             & $\texttt{row} (b_{n-1}, \; b_{n-2},\dots, \; a +b_{0})$\\
         1 & sum-of-rows $\dagger$ & $\texttt{sum}(\shrow{a}, \; \shrow{b})$ & $\texttt{row} (\texttt{sum}(a_{n-1}, \; b_{n-1}), \; \dots  \texttt{sum}(a_{0}, \; b_{0}))$\\
         1 & sum-of-bits $\dagger$ & $\texttt{sum}(a, b, c, ...)$ & $a+b+c\; ...$\\
         1 & row-of-rows $\dagger$ & $\texttt{row} (\shrow{a}, \; \shrow{b})$ & 
         $\texttt{row} (a_{n-1}, \; \texttt{sum}(\texttt{row}(a_{n-2}, \dots, a_0), \; \texttt{row}(b_{n-1} \dots b_{1})), \; b_0)$\\
         1 & repeated-bit    & $\rrow{a}$                & $\texttt{sum}(\rrow{1},\;\bar{a})$\\
         \hline 
         2 & Half-Adder Sum    & $HA_s(a,b)$               & $a \oplus b$ \\
         2 & Half-Adder Carry  & $HA_c(a,b)$               & $a \& b$ \\
         2 & Full-Adder Sum    & $FA_s(a,b,c)$             & $a \oplus b \oplus c$ \\
         2 & Full-Adder Carry  & $FA_c(a,b,c)$             & $a\,\&\,b\; \| \; c\&(a\oplus b)$ \\
         2 & Sop Xor           & $a\oplus b $              & $(a\,\&\,\bar{b})\, \|\, (\bar{a}\,\&\,b)$ \\
         2 & De Morgan And     & $\overline{a \& b}$       & $\bar{a} \| \bar{b}$ \\
         2 & De Morgan Or      & $\overline{a \| b}$       & $\bar{a} \& \bar{b}$ \\
         2 & Distrib. And Or  & $a \& (b \| c)$       & $(a\&b) \| (a\&c)$ \\
         2 & Distrib. And Xor & $a \& (b \oplus c)$   & $(a\&b) \oplus (a\&c)$ \\
         2 & Xor And          & $a \oplus(a \& b)$   & $a\&\overline{b}$ \\
         2 & Or Not And       & $\overline{a} \| (a \& b)$ & $\overline{a}\| b$ \\
         \hline
         

         
         
         
        \rowcolor{white} Pre & divide-and-conquer $\dagger$ & $\shrow{a} * \shrow{b}$ & 
        $\texttt{sum}(\texttt{row}(a_{n-1} \dots a_{n/2}) * \texttt{row}(b_{n-1} \dots b_{n/2})\ll n,$\\ 
        \rowcolor{white} & ($n$ even)&&$\hspace{1.7em}\texttt{row}(a_{n-1} \dots a_{n/2}) * \texttt{row}(b_{n/2-1} \dots b_{0})\ll n/2,$\\
        \rowcolor{white} & &&\hspace{2.2em}$\texttt{row}(a_{n/2-1} \dots a_{0})) * \texttt{row}(b_{n-1} \dots b_{n/2})\ll n/2,$\\
        \rowcolor{white} & &&\hspace{-1.5em}$\texttt{row}(a_{n/2-1} \dots a_{0}) * \texttt{row}(b_{n/2-1} \dots b_{0}))$\\ 
         
    \hline
         
    \end{tabular}
\end{table*}




An e-graph is grown via rewriting. OptiMult includes two rewrite sets. The first rewrite set encodes how, at a higher level of abstraction, an input AND array can be reduced to a \texttt{row} of single bit expressions, essentially performing array reduction and carry-propagation at the compressor cell level of abstraction. The second set of rewrites decomposes the compressor cell operators into a gate-level representation and performs gate-level rewriting. The rewrites are described in Table~\ref{tab:rewrites} and are separated according to the phase in which they are applied. 

The first set contains rewrites such as ``Place Full/Half-Adder'' that apply a compressor cell reducing the addition of three/two bits to a \texttt{row} of two bits. The ``add-same'' rewrite detects the opportunity to exploit common signals, essential for squarer circuit optimization. The ``sum-of-rows'' rewrite matches a sum of rows representation (Figure~\ref{tab:sum_of_rows}), and converts it into the row of sums representation (Figure~\ref{tab:row_of_sums}). The ``row-of-rows'' rewrite performs the inverse. To enable compressor cell placement, the ``sum-of-bits'' rewrite converts a \texttt{sum} that only consists of single bit expressions into an adder chain. This rewrite set encodes the transformations to reduce an AND array to a single bit expression for each output bit. The ``repeated-bit'' rewrite performs a multiplier optimization to collect repeated sign extension bits into a row of constant bits plus a correction. These constants can then be automatically combined with rows above or below.

The second set contains the ``Full/Half-Adder Sum/Carry'' rewrites that transform the higher abstraction level operators into gate-level implementations. This expands the exploration space as it greatly increases the number of operators in the e-graph. The set also contains the standard rules of commutativity, associativity and distributivity for the Boolean operators, AND, OR and XOR. The standard Boolean rewrites are omitted from Table~\ref{tab:rewrites}. Other Boolean rewrites are also included such as De-Morgan's Laws and XOR decomposition, plus rewrites learnt from analyzing the generated expressions. These rewrites encode Boolean optimizations reducing the expressions for each output bit to a delay optimized form. 

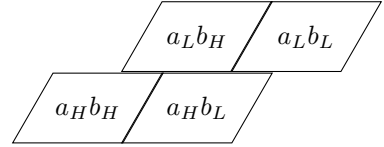
\begin{figure}
    \centering
\begin{tikzpicture}
   \node[trapezium, draw, trapezium left angle=60, trapezium right angle=120, minimum width = 2cm]
         at (0,0) {};
    \node[trapezium, draw, trapezium left angle=60, trapezium right angle=120, minimum width = 2cm]
         at (1.45,0) {};
    \node[trapezium, draw, trapezium left angle=60, trapezium right angle=120, minimum width = 2cm]
         at (2.9,0.95) {};
    \node[trapezium, draw, trapezium left angle=60, trapezium right angle=120, minimum width = 2cm]
         at (1.45,0.95) {};
    \node [] at (1.45,0) {$a_H b_L$};
    \node [] at (0,0) {$a_H b_H$};
    \node [] at (2.9,0.95) {$a_L b_L$};
    \node [] at (1.45,0.95) {$a_L b_H$};
    \end{tikzpicture}
    \caption{The ``divide-and-conquer'' rewrite represented in an AND array, that decomposes a large multiplier $a* b$, into four smaller multipliers by splitting $a=\{a_H,a_L\}$ and $b=\{b_H,b_L\}$.}
    \label{fig:divide_and_conq}
\end{figure}

E-graph rewriting of gate-level expressions suffers from scalability issues. To optimize higher bitwidth multipliers, Table~\ref{tab:rewrites} includes the ``divide-and-conquer'' rewrite, which expresses larger multipliers as a sum of smaller multipliers (Figure~\ref{fig:divide_and_conq}). OptiMult can then re-use prior optimization results, to map the smaller multipliers directly to efficient representations. The e-graph then only needs to compute a reduction of four rows rather than a larger array. In this way, larger multipliers can be generated from efficient smaller multipliers. Without this technique even an efficient 8-bit square design is unreachable.

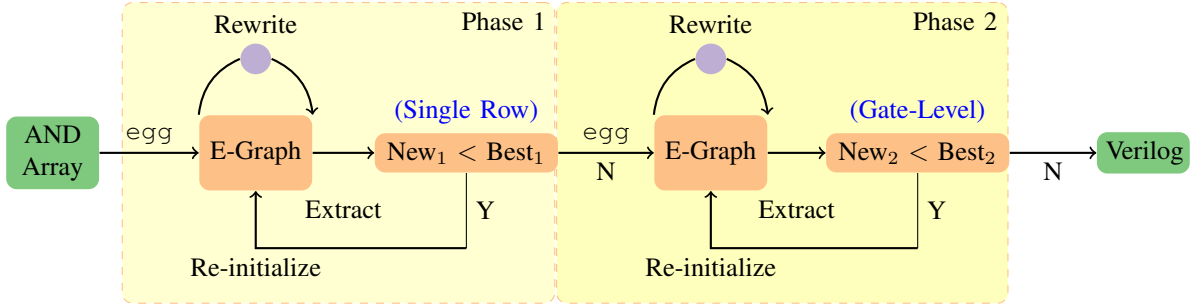
\begin{figure*}
    \centering
    \begin{tikzpicture}


\node [shape=rectangle, fill=scyellow!45,minimum width=5.75cm, minimum height=4cm,draw=scorange,dashed,rounded corners] at (-0.2,0) (phase2) {};

\node [shape=rectangle, fill=scyellow!65,minimum width=6cm, minimum height=4cm,draw=scorange,dashed,rounded corners] at (5.7,0) (phase3) {};

\node[] at (2,1.75) (phase2) {Phase 1};
\node[] at (8,1.75) (phase3) {Phase 2};

\node[] at (1.5,0.55) (phase1_tgt) {\textcolor{blue}{(Single Row)}};
\node[] at (7.5,0.55) (phase1_tgt) {\textcolor{blue}{(Gate-Level)}};

\node [shape=rectangle, fill=scgreen, text width = 1cm, text centered,rounded corners] at (-4,0) (and) {AND Array};

\node [shape=rectangle, fill=scorange, minimum width=1.5cm, minimum height=1cm,rounded corners] at (-1.3,0) (egraph2) {E-Graph};
\node [shape=circle, fill=scpurple, text width = 0.1cm, text centered,label={Rewrite}] at (-1.3,1.25) (rewrites2) {};
\node [shape=rectangle, fill=scorange, text width = 2.2cm, text centered, rounded corners] at (1.5,0) (single_row) {New$_1$ $<$ Best$_1$};

\node [shape=rectangle, fill=scorange, minimum width=1.5cm, minimum height=1cm,rounded corners] at (4.75,0) (egraph3) {E-Graph};
\node [shape=circle, fill=scpurple, text width = 0.1cm, text centered,label={Rewrite}] at (4.75,1.25) (rewrites3) {};
\node [shape=rectangle, fill=scorange, text width = 2.2cm, text centered,rounded corners] at (7.5,0) (opt) {New$_2$ $<$ Best$_2$};

\node [shape=rectangle, fill=scgreen, text width = 1cm, text centered,rounded corners] at (10.5,0) (verilog) {Verilog};

\draw [thick,->] (and) -> (egraph2) node [above,midway] {\texttt{egg}};

\draw [thick,->] (egraph2) -> (single_row) node [below=0.5cm,midway] {Extract};

\draw [thick,->] (single_row) -> (egraph3) node [above,midway] {\texttt{egg}} node [below,midway] {N};
\draw [thick,->] (egraph3) -> (opt) node [below=0.5cm,midway] {Extract};
\draw [thick,->] (opt) -> (verilog) node [below,midway] {N};


\path [->, thick] (rewrites2.east) edge[bend left] (egraph2.north east);
\draw [-, thick] (egraph2.north west) edge[bend left] (rewrites2.west);

\path [->, thick] (rewrites3.east) edge[bend left] (egraph3.north east);
\draw [-, thick] (egraph3.north west) edge[bend left] (rewrites3.west);

\coordinate (x) at ([yshift=-1cm]single_row.south);
\coordinate (y) at (single_row.south |- x);
\draw (single_row.south) -- (y) (x) node [midway,right] {Y};
\draw [->, thick] (y)(x) -| (egraph2.south) node [midway,below] {Re-initialize};

\coordinate (x) at ([yshift=-1cm]opt.south);
\coordinate (y) at (opt.south |- x);
\draw (opt.south) -- (y) (x) node [midway,right] {Y};
\draw [->, thick] (y)(x) -| (egraph3.south) node [midway,below] {Re-initialize};


\end{tikzpicture}
    \caption{An overview of the OptiMult tool flow. Each phase has an associated rewrite set and delay cost model, targeting the structure written in blue. After each e-graph extraction the delay cost of the new design is compared against the previous best. While the design continues to improve in terms of the estimated delay cost, the e-graph is re-initialized with the new design to repeat the current phase.}
    \label{fig:flow_diag}
\end{figure*}

\subsection{Rewrite Scheduling and Extraction} \label{sec:schedule}

Figure~\ref{fig:flow_diag} provides an overview of the OptiMult flow, where optimization is separated into two phases. Each phase has a set of rewrites associated with it and OptiMult grows an e-graph in each phase starting from the expression extracted in the previous phase. The e-graph is grown until a user-defined rewrite iteration or node limit is reached. Each phase has an associated cost model that drives the extraction towards a particular target structure, returning a single implementation after extraction. Each phase operates at a decreasing abstraction level, going from compressor cell placement in the first phase to gate-level optimization in the second phase. 

Both cost models share a common base for well-defined operators to estimate the delay of the multiplier designs. The base model assigns one to single gate operators and a total delay in gates to more complex operators like compressor cells. The cost models for each phase primarily differ in how they assign costs to the \texttt{row} and \texttt{sum} operators. The phase one cost model targets a single row of individual bit expressions, assigning high cost to any \texttt{sum} and any \texttt{row}s that are not comprised of single bit expressions but does not impose any bias against high-level operators such as the Full/Half-Adder Carry and Sum operators, discussed in Section~\ref{sec:representation}. In phase two, OptiMult assumes it receives a single row representation, so now targets an optimized gate-level implementation, assigning high cost to the compressor cell operators that are automatically rewritten as Boolean expressions. 

All cost models assume that every operator has a fan-out of one, because correctly evaluating the cost of common sub-expressions in an e-graph is computationally expensive~\cite{Coward2022AutomaticE-Graphs,Wang2020SPORES:Algebra}. This means that the e-graph essentially optimizes the expression to generate each output bit independently but shares the optimization effort across all bits. The impact of this assumption on the results shall be analyzed in Section~\ref{sec:arch_res}.

To further optimize, OptiMult repeatedly runs each phase until the extracted cost stops decreasing. The phasing of e-graph optimization limits the exponential growth of the e-graph as only the rewrites needed to progress at each stage are applied. Complementing the distinct phases with the re-initialization approach limits the exponential growth of the e-graph, as the number of equivalent gate-level multipliers grows exponentially with the multiplier bitwidth. Further scalability gains can be accessed by composing optimized small blocks to produce larger operators, a common approach in arithmetic component design. For example, as a preprocessing step, the ``divide-and-conquer'' rewrite can be used to decompose higher-bitwidth multipliers into sums of smaller multipliers. The final output design is automatically translated to synthesizable Verilog.

\section{Results}\label{sec:results}
We first analyze the OptiMult generated 3-bit implementation and squarer implementations to understand any architectural novelties discovered by the e-graph. We then compare the unsigned multipliers and squarers generated by OptiMult against those implemented by a commercial logic synthesis tool, which we call the specification, for a TSMC 5nm cell library. All generated multipliers are formally verified using a commercial equivalence checker. 

\subsection{Architectural Results} \label{sec:arch_res}
\begin{figure}
    \centering
    \includegraphics[scale=0.75]{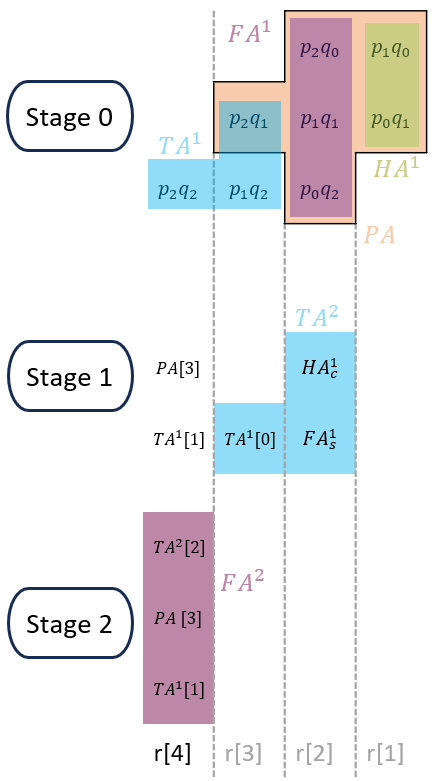}
    \caption{OptiMult discovered circuit computing bit \texttt{r[4]} for a 3-bit multiplier, \texttt{r = p * q}. The circuit deploys several compressor variants, Full-Adders (FA), Half-Adders (HA), Triangle-Adders (TA) and P-Adders (PA).}
    \label{fig:3-bit-mult}
\end{figure}

In this section we analyze the OptiMult generated 3-bit multiplier to explain the performance exhibited in the synthesis results. Firstly, as discussed in Section~\ref{sec:schedule}, each output bit is independently optimized meaning the generated circuit does not exploit common intermediate signals. In Figure~\ref{fig:3-bit-mult} we show how the OptiMult generated multiplier constructs a single bit of the output. 

The generated circuit does not respect a well-defined structure and discovers a number of non-standard compressor cells that it deploys in addition to the standard Full- and Half-Adders. For example, the generated multiplier uses a three input compressor (blue) producing three output bits, which we have called a Triangle Adder (TA), to exploit common terms that it can factorize. 
\begin{align*}
    TA(p_2q_1, p_1q_2, p_2q_2) = &\; 2* p_2q_2 + p_2q_1 + p_1q_2\\                        
    TA[2] = &\;  p_2q_1p_1q_2\\
    TA[1] = &\; p_2q_2\overline{p_1q_1}\\
    TA[0] =&\; p_2q_1\oplus p_1q_2
\end{align*}
The same TA can equivalently be computed as a sequence of Half-Adders,
\begin{align*}
    TA[2] =&\;HA_c(HA_c(p_2q_1, p_1q_2), p_2q_2),\\ 
    TA[1] =&\;HA_s(HA_c(p_2q_1, p_1q_2), p_2q_2),\\ 
    TA[0] =&\;HA_s(p_2q_1, p_1q_2).
\end{align*}
The Half-Adder circuit does not exploit correlation amongst the bits, namely the most significant bit would be computed as $p_2q_1p_1q_2p_2q_2$. This could be simplified by gate-level rewriting. As a result, the Half-Adder circuit generates the most significant two bits with a delay of three gates, whilst the TA computes the same bits with only a two gate delay. 

The second $TA$, $TA^2$, demonstrates how OptiMult balances delay. 
\[
    TA^2[2] = \overbrace{(TA^1[0] \cdot HA_c^1)}^{\textrm{3 gates}}\cdot \overbrace{FA_s^1}^{\textrm{3 gates}}
\]
OptiMult factorizes the expression in this way because bits $TA^1[0]$ and $HA_c^1$ arrive one gate earlier than bit $FA_s^1$.

Figure~\ref{fig:3-bit-mult} also deploys a six input compressor with a 4-bit result, which we call a P-adder.
\begin{align*}
    &PA(p_1q_0, p_0q_1, p_2q_0, p_1q_1, p_0q_2, p_2q_1) =\\ 
    &4*p_2q_1 + 2* (p_2q_0 + p_1q_1 + p_0q_2) + p_1q_0 + p_0q_1
\end{align*}
In Figure 4, OptiMult only uses the fourth output bit from the P-adder, however the other bits can also be computed efficiently.
\begin{align*}
    PA[3] = &\; K\; \cdot\; FA_c(p_2q_0,\,p_1q_1,\, p_0q_2)\\
    PA[2] = &\; K \oplus  FA_c(p_2q_0,\,p_1q_1,\, p_0q_2)\\
    PA[1] = &\; (p_1q_1\overline{p_0q_0}) \oplus (p_2q_0 \oplus p_0q_2)      \\
    PA[0] = &\; p_1q_0\oplus p_0q_1\\
    \textrm{where, }K =&\; (p_2 \oplus ((p_2q_0 \oplus p_1 \oplus p_0q_2)p_0p_1q_0))q_1
\end{align*}
To compute bit $PA[1]$ a TA is used that allows that bit to be computed using only seven gates with a three gate delay. An alternative is to deploy first a Half-Adder and Full-Adder, followed by another Half-Adder. Such a circuit computes the same bit using nine gates with a four gate delay:
\begin{equation*}
    (p_1q_0p_0q_1) \oplus p_2q_0 \oplus p_1q_1 \oplus p_0q_2.
\end{equation*}

In the $PA$ expressions a Full-Adder is deployed, utilizing the carry-out of $FA^1$ in Figure~\ref{fig:3-bit-mult}. This may result in a sub-optimal circuit delay. By hand we can construct an alternative and lower delay circuit to compute bit $P[3]$.
\begin{equation*}
    PA[3] = p_2q_1(p_0q_0q_2 \| p_1(q_0 \oplus p_0q_2))
\end{equation*}
We hypothesize that OptiMult is driven towards the original design by the phase one compressor cell placement. 

The overall critical path of Figure~\ref{fig:3-bit-mult} is one PA and one FA Sum. The fastest circuit that could be constructed using FAs and HAs only would have a delay of one FA Carry and three HAs. Note that the expression to construct the PA has a gate delay comparable to two FAs. 

The squarer circuits exploit common signals throughout to substantially reduce the gate count. For example, OptiMult reduces the 3-bit squarer to the following, where $o_i$ denotes the $i$\textsuperscript{th} output bit:
\begin{align*}
o_5 =&\; p_2 p_1                       & o_2 =&\; p_1 \overline{p_0} \\
o_4 =&\; ((p_0 \| \overline{p_1})p_2) & o_1 =&\; 0 \\
o_3 =&\; p_0 (p_2 \oplus p_1)          & o_0 =&\; p_0.
\end{align*}
The squarer optimization exploits both column-wise and row-wise transformations, detecting identical bits in a column and carrying them forward through the \texttt{row}. This implementation is the same as that obtained via traditional gate-level transformations, which demonstrates that OptiMult generalizes known techniques. The dual representation described in Section~\ref{sec:representation}, allows OptiMult to more efficiently reach such implementations ensuring that optimized higher bitwidth squarers are reachable.  

In Section~\ref{sec:schedule} we discussed a phased e-graph optimization approach to address scalability. If we combine phases one and two into a single phase, OptiMult takes 42 iterations to converge for the 3-bit square. Using the two phase approach shown in Figure~\ref{fig:flow_diag}, OptiMult only requires 31 iterations for the 3-bit square. This problem becomes worse for larger bitwidth designs, such that separating the optimization into phases is essential to reach optimized architectures.

\subsection{Synthesis Results} \label{sec:synth_res}
For a range of bitwidths we synthesize the gate-level OptiMult generated unsigned multipliers and compare them against the specification components instantiated by logic synthesis when writing the $*$ operator. For each architecture, we synthesize the competing designs at a zero delay target to obtain a minimum achievable latency. Table~\ref{tab:synth_data} contains the synthesis results. 

We observe that for small bitwidth squarer circuits, the design instantiated from $p*p$ by logic synthesis and the OptiMult generated gate-level implementation appear identical. For larger bitwidths, we observe a significant latency reduction from the OptiMult generated squarers. When synthesizing the specification the synthesis tool clearly exploits the fact that it is a square as the 5-bit square specification is significantly faster than the general 5-bit multiplier. As the bitwidth increases the number of possible factorizations of each output bit grows, meaning the search space is substantially larger. OptiMult is able to explore a large number of these factorizations and iteratively refine the expression for each output bit. The result is a highly optimized Boolean expression for each output bit. OptiMult exploits common bits throughout allowing it to discover efficient factorizations. 

\begin{figure}
    \centering
    \begin{tikzpicture}[]
	\begin{axis}[%
		xlabel=Delay (ps),ylabel=Area ($\mu m^2$)]

\addplot[
    color=red,
    mark=square,
    ]
    coordinates {
(71.9,6.993630)
(72,6.104700)
(80,4.091220)
(90,3.234420)
(100,3.352230)
(110,3.384360)
};
\addlegendentry{Spec}

\addplot[
    color=blue,
    mark=square,
    ]
    coordinates {
(51.2,7.646940)
(72,4.026960)
(80,3.245130)
(90,3.020220)
(100,3.052350)
(110,2.848860)
    };
    \addlegendentry{OptiMult}

    \end{axis}
\end{tikzpicture}
    \caption{Area delay profile for the competing 6-bit squarer implementations.}
    \label{fig:area-delay}
\end{figure}
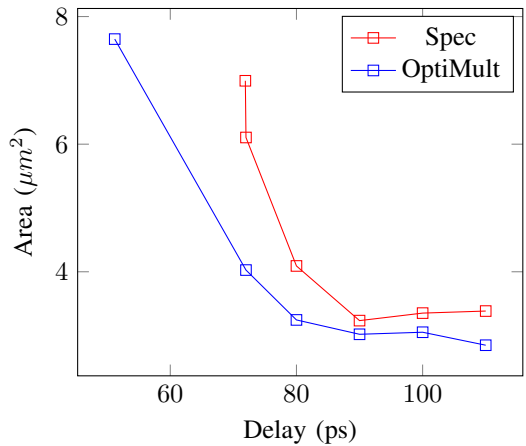

In Figure~\ref{fig:area-delay} we plot an area-delay profile comparing the synthesis instantiated squarer against the OptiMult generated squarer. The plot shows that the OptiMult generated squarer is not only faster than the specification, but is up to 34\% smaller. 

For the general multiplier circuits, we see the opposite trend. For the lower bitwidth implementations, the OptiMult generated multipliers demonstrate latency reductions whilst for larger bitwidths the generated multiplier exhibits near identical performance. For the smaller bitwidth multipliers, OptiMult is able to generate an efficient reduction for each output bit discovering generalized compressors cells that exploit common signals. The 5-bit general multiplier hits an e-graph scalability limit, meaning that OptiMult is unable to progress this design to an efficient implementation. For example, the 5-bit multiplier can be up to 45\% larger than the specification. The carry-bit computations for higher-order bits become exceedingly complex as OptiMult is unable to reduce the Boolean expressions for the output bits. Traditional multiplication architectures typically scale with more structure.

\begin{table}[t]
    \centering
    \begin{tabular}{|c|r|r|r|}
        \hline
         Design     & Spec (ps) & OptiMult (ps) & Change (\%)\\
        \hline
         3-bit Sqr.   &  9.5    &  9.5   & 0.0     \\ 
         4-bit Sqr.   & 13.4    & 13.4   & 0.0     \\ 
         5-bit Sqr.   & 54.6    & 29.3   & -46.2   \\ 
         6-bit Sqr.   & 71.9    & 51.2   & -28.8   \\ 
         8-bit Sqr.   & 93.6    & 73.8   & -21.2   \\ 
         \hdashline
         3-bit Mult   & 38.6    & 35.3   & -8.5   \\ 
         4-bit Mult   & 65.2    & 62.8   & -3.7   \\ 
         5-bit Mult   & 91.5    & 89.3   & -2.3    \\ 
         \hline
    \end{tabular}
    \caption{Synthesis results comparing unsigned multipliers and squarers instantiated from Verilog $*$ operator against the gate-level OptiMult generated implementations. The designs are synthesized at a zero delay target to obtain a minimum achievable latency.}
    \label{tab:synth_data}
\end{table}

\section{Conclusion}
This paper has described how traditional multiplier optimizations can be encoded as local equivalence preserving transformations. We explored how these rewrites can be combined using an e-graph, removing preconceived ideas about optimality. The e-graph approach is able to efficiently explore the large design space of equivalent multiplier implementations. The generated multipliers exhibit unusual patterns that exploit common signals and do not resemble traditional compressor cell structures. The resulting circuits are shown to be competitive with the multipliers instantiated by a commercial logic synthesis tool, which deploys state-of-the-art components.

Future work will expand the rewrite set to include more generalized compressor cells and to explore how OptiMult manipulates these structures. For example, the implementation of the $(7:3]$ counter described in~\cite{Ercegovac2004DigitalArithmetic}, is a complex Boolean factorization problem in itself, without taking into account the wider context. In addition, we will investigate common operator encoding schemes for multipliers such as Booth Radix-4~\cite{Ercegovac2004DigitalArithmetic} and alternative approaches such as left-to-right carry free multipliers~\cite{Ercegovac1990FastAddition}.




\ifCLASSOPTIONcompsoc
  \section*{Acknowledgments}
\else
  \section*{Acknowledgment}
\fi

The authors would like to thank Emiliano Morini who suggested the tool's name, OptiMult.

\bibliographystyle{IEEEtran}
\bibliography{IEEEabrv,references}

\end{document}